\documentclass[aps,prd,preprint,tightenlines,floatfix]{revtex4}
\usepackage{graphicx}

\newcommand{\nc}{\newcommand}

\nc{\beq}{\begin{equation}}
\nc{\eeq}{\end{equation}}
\nc{\bea}{\begin{eqnarray}}
\nc{\eea}{\end{eqnarray}}
\nc{\n}{\nonumber \\}
\nc{\K}{27 \, \textrm{mK} \, \sqrt{ \frac{1+z}{10} }  \; }
\nc{ \X} {$x_{\rm ion}$}

\begin{document}

\date{June 11, 2010}
\title{Distinguishing standard reionization from dark matter models}
\author{Aravind Natarajan}
\email{anat@andrew.cmu.edu}
\affiliation{McWilliams Center for Cosmology, Carnegie Mellon University, Department of Physics, 5000 Forbes Ave., Pittsburgh PA 15213, USA}

\author{Dominik J. Schwarz}
\email{dschwarz@physik.uni-bielefeld.de}
\affiliation{Fakult\"{a}t f\"{u}r Physik, Universit\"{a}t Bielefeld, Universit\"{a}tsstra\ss e 25,  33615 Bielefeld, Germany}

\begin{abstract} 
The Wilkinson Microwave Anisotropy Probe (WMAP) experiment has detected reionization at the $5.5 \sigma$ level and has reported a mean optical depth of $0.088 \pm 0.015$. A powerful probe of reionization is the large-angle $EE$ polarization power spectrum, which is now (since the first five years of data from WMAP) cosmic variance limited for $2\le l \le6$.  Here we consider partial reionization caused by WIMP dark matter annihilation, and calculate the expected polarization power spectrum. We compare the dark matter models with a standard 2-step reionization theory, and examine whether the models may be distinguished using current, and future CMB observations.  We consider dark matter annihilation at intermediate redshifts ($z<60$) due to halos, as well as 
annihilation at higher redshifts due to free particles.  In order to study the effect  of 
high redshift dark matter annihilation on CMB power spectra, it is essential to include the 
contribution of residual electrons (left over from recombination) to the ionization history.
Dark matter halos at redshifts $z<60$ 
influence the low multipoles $l<20$ in the $EE$  power spectrum, while the annihilation of 
free particle dark matter at high redshifts $z>100$ mainly affects multipoles $l>10$. 
\end{abstract}

\maketitle

\section{Introduction}

Studies of quasar spectra indicate that the Universe is highly ionized up to a redshift $z\sim6$.  In the standard approach, early stars, early galaxies, quasars, etc.~are thought to be the sources responsible for reionizing the Universe. The free electrons that exist after reionization scatter microwave photons, resulting in an increase in the degree of polarization of the cosmic microwave background (CMB) radiation. The Wilkinson Microwave Anisotropy Probe (WMAP) has confirmed the reionization of the Universe and has computed the optical depth due to scattering of CMB photons with free electrons
\cite{wmap_first_year_1, wmap_three_year_1, wmap_five_year_1, wmap_five_year_2, wmap_7year_1}.

There is considerable uncertainty regarding the extent to which early stars reionize the Universe. The epoch of formation of the first stars, their masses, their abundance, etc are model dependent factors that influence the ionization efficiency. There are also questions regarding the nature of the first stars. In some models with weakly interacting massive particles (WIMPs) being the dark matter, the earliest generation of stars were dark stars \cite{darkstars} characterized by a low surface temperature, and very high mass. In these theories, it is unclear how the next generation of stars formed, and whether they were able to partially reionize the Universe. Here, we will consider dark matter annihilation to be a source of reionization.

Several authors have studied the impact of particle decay and annihilation on reionization \cite{ion}. Annihilating dark matter with mass $m_\chi \sim 10 - 100$ GeV may be detectable by future CMB polarization measurements  \cite{pad}. More recently, there has been an interest in constraining the properties of dark matter models using the WMAP measured optical depth. In previous work \cite{arvi1,arvi2}, we studied dark matter halos with low WIMP masses, with the restriction that they be consistent with the measured optical depth. Heavier dark matter particles $m_\chi \gtrsim 100$ GeV 
in halos could reionize the Universe for favorable particle physics parameters \cite{hooper}, and could possibly account for the positron excess in high energy cosmic rays found by ATIC 
\cite{Chang:2008zzr} and PAMELA \cite{Adriani:2008zr}. 
Reionization by annihilating dark matter for a variety of particle and cosmological parameters was presented in \cite{more}; reference \cite{more2} studies the allowed regions in the dark matter mass-cross section plane. More recently, \cite{new} have analyzed a combination of recent CMB data sets
for evidence of dark matter annihilation. In this article, we investigate whether it is possible to distinguish standard reionization theories from more exotic dark matter scenarios, using CMB observations. 

Scattering of free electrons by CMB photons causes additional polarization at scales $\sim$ the horizon at reionization, resulting in excess power at low multipoles compared to a Universe that is not reionized. Let us denote the WMAP measured optical depth as $\tau$. We note that this is generally different from $\tau(z)$, the total optical depth up to redshift $z$. This is because the WMAP experiment may not be sensitive to large $z$, and also because of the effect of residual electrons included in $\tau(z)$.  The large angle $EE$ polarization $\propto \tau(z)^2$ is an excellent probe of reionization, but is seen at high significance by WMAP only for $l<10$ \cite{wmap_five_year_1,wmap_7year_1}. 
With 1 year of data, the WMAP experiment detected a reionization signal in the $TE$ polarization power spectrum, implying an optical depth $\tau = 0.17 \pm 0.04$ \cite{wmap_first_year_1}. This value was shown to be too large by the WMAP 3-year analysis \cite{wmap_three_year_1}. With 3 years of data, WMAP obtained a value for $\tau = 0.10 \pm 0.03$ using $EE$ data alone, and $\tau = 0.09 \pm 0.03$ using $EE$, $TE$, and $TT$ data \cite{wmap_first_year_1}. The 3-year measurement of $\tau$ has been confirmed by the 5-year analysis \cite{wmap_five_year_1, wmap_five_year_2}. The 5-year WMAP measured $EE$ power spectrum is largely cosmic variance limited at multipoles $2\le l \le6$ \cite{wmap_five_year_1}. The WMAP 5-year analysis 
provides a mean value $\tau=0.087 \pm 0.017$ using all WMAP data \cite{wmap_five_year_2}. With BAO and SN data included, the measured optical depth is $\tau = 0.084 \pm 0.016$ \cite{params}.  Assuming an instantaneous reionization model, WMAP reports a reionization epoch $z_\ast = 11 \pm 1.4$ using WMAP data alone, and $z_\ast = 10.8 \pm 1.4$ using WMAP+BAO+SN data \cite{params}.  These results are in excellent agreement with the recent 7-year result $\tau = 0.088 \pm 0.015$ for sudden reionization and $\tau = 0.087\pm 0.015$ for a fit which allowed a varying width of a one step reionization scenario. The low $l$ polarization data (EE) alone, implies a detection of reionization at the $5.5\sigma$ level \cite{wmap_7year}. 

In Section II, we provide a brief discussion of dark matter annihilation and the inverse compton scattering process. We compute the energy absorbed by gas at a given redshift, and the expected ionization and temperature histories. In Section III, we consider dark matter models with different particle masses and concentration parameters, and calculate the expected polarization power spectra using the {\scriptsize CAMB} program \cite{camb} with appropriate modifications. We first consider dark matter annihilation by free 
(gravitationally unbound) particles at high redshifts $z>100$ with and without residual electrons. Ionization in this regime directly probes the particle physics properties of the model. We then consider ionization at all redshifts, including WIMP annihilation in halos, and compare with a standard 2-step reionization model.  Finally, we present our conclusions. We used the parameters $\Omega = 1$, $\Omega_{\rm b}h^2 = 0.022$, $\Omega_{\rm dm}h^2 = 0.12$, $h=0.7$, $A_{\rm s} = 2.3\times 10^{-9}$, $n_{\rm s} = 1$, $w=-1$, $T_{\rm cmb} = 2.726$ K to make our plots.

\section{Particle annihilation in halos.}

The number of dark matter annihilations per unit volume and per unit time, at a redshift $z$ is given by:
\beq
\frac{dN_{\rm ann}}{dtdV} (z) = \frac{\langle \sigma_{\rm a} v \rangle}{2 m^2_\chi} \, \rho^2_\chi (z).
\eeq
$\langle \sigma_{\rm a} v \rangle$ is the annihilation cross section times the relative velocity of the particles, averaged over the WIMP velocity distribution. $m_\chi$ is the particle mass, and $\rho_\chi(z)$ is the dark matter density at $z$. We have assumed that two WIMPs participate in an annihilation process producing particles with total energy $2 m_\chi c^2$ per annihilation. Before the epoch of halo formation, $\rho^2_\chi$ takes the simple form:
\beq
\rho^2_{\chi, \rm free} (z) = (1+z)^6 \, \rho^2_{\rm c} \, \Omega^2_{\rm dm},
\eeq
where $\rho_{\rm c}$ is the critical density, and $\Omega_{\rm dm}$ is the dark matter fraction today. 

As the annihilation rate is proportional to the square of the WIMP density, it is enhanced by the formation 
of structure in the Universe. Hierarchical structure formation starts at a redshift $z\simeq 60$ 
\cite{ghs, Green:2005fa}. The boost in density due to clustering of WIMPs is partially countered by the small halo volume. Nevertheless, $\rho_{\chi, \rm{halo}}$  eventually becomes much larger than $\rho_{\chi,\rm{free}}$ (unless the concentration parameter is very small). $\rho_{\chi, \rm halo}$ is given by:
\beq
\rho^2_{\chi, \rm halo} (z) = (1+z)^3  \, \int_{\rm M_{\rm min}} dM \, \frac{dn_{\rm halo}}{dM} (M,z) \left [ \int_0^{r_{\rm{200}}} dr \, 4 \pi r^2 \rho^2_{\rm h}(r) \right ] (M,z).
\label{formula_halo}
\eeq
$n_{\rm halo}$ is the comoving number density of halos. The volume integral over the halo is a function of both halo mass $M$ and redshift $z$. Here we have assumed that halos are truncated at $r_{\rm 200}$, the radius at which the mean density enclosed equals 200 times the background density at the time of halo formation. $\rho_{\rm h}(r)$ is the halo density at a distance $r$ from the halo center. Eq. (\ref{formula_halo}) ignores halo-halo interactions, as well as interactions between free dark matter and 
dark matter in halos. 
 
\subsection{Ionization and heating.}

WIMP particle annihilation generally results in the release of photons, charged particles, and neutrinos. Let $dN_\gamma / dE_\gamma(z)$ be the number of photons produced per unit energy per annihilation at $z$. $dN_\gamma/dE_\gamma$ includes both prompt photons released by WIMP annihilation, as well as lower energy photons produced as a result of processes such as inverse compton scattering (ICS henceforth), bremsstrahlung, etc. The energy absorbed at a redshift $z$ per gas atom $\xi(z)$ can be calculated as \cite{arvi2}:
\beq
\xi(z) = \int_\infty^z \frac{-dz'}{(1+z') H(z')} \, \left( \frac{1+z}{1+z'} \right )^3 \, \left( \frac{dN_{\rm ann}}{dt dV} \right ) (z') \;  \int_{E_{\rm 1}} ^{E_{\rm 2}}  dE'_\gamma \, E'_\gamma \; \frac{dN_\gamma}{dE'_\gamma} (E'_\gamma) \; e^{-\kappa(z',z; E'_\gamma)} \; \left[ c \sigma(E'_\gamma) \right ].
\label{energy_equation}
\eeq
In Eq. (\ref{energy_equation}), we expressed the time variable $dt$ in terms of $dz$ using the relation $dz = -dt (1+z) H(z)$, where $H(z)$ is the Hubble parameter at redshift $z$. The cubic term accounts for the expansion of the Universe in the time it takes for the photons to travel from $z'$ to $z$. $E'_\gamma = E_\gamma (1+z) / (1+z')$, which accounts for the redshifting of photon energy. We have assumed that the photons are produced at the redshift of WIMP annihilation $z'$. This is exact for prompt photons, and a good approximation for ICS photons produced by charged particles interacting with the CMB. $E_1$ and $E_2$ are suitably redshifted energies. The last term in square brackets accounts for scattering with gas atoms at $z$. $c$ is the speed of light, while $\sigma$ is the cross section.  The exponential term accounts for the scattering of photons as they travel from $z'$ to $z$. $\kappa(E_\gamma)$ is the optical depth for photons of energy $E_\gamma$, given by the expression
\beq
\kappa(z',z; E_\gamma) = \int_{z'}^{z} \frac{- dz'' }{(1+z'') H(z'')} \; c \, n(z'') \sigma(E_\gamma).
\label{kappa}
\eeq

Let $\eta_{\rm ion}(z)$ and $\eta_{\rm heat}(z)$ be the fractions of energy that go into ionization and heating respectively. These fractions have been calculated by \cite{shull_vansteen, kanzaki_energy, furl0,valdes}.  For simplicity, we set $\eta_{\rm ion} = \eta_{\rm heat} = 1/3$, with the remaining contributing to collisional excitations and the Ly-$\alpha$ background. The ionized fraction $x_{\rm ion}(z)$ and the gas temperature $T(z)$ may be calculated by solving together, the two equations:
\bea
-(1+z) H(z) \frac{dx_{\rm ion}(z)}{dz}  &=& \mu \left [1-x_{\rm ion}(z) \right ] \eta_{\rm ion} (z)  \xi(z)   - n(z) x^2_{\rm ion}(z) \alpha(z)  \n
-(1+z) H(z) \frac{dT(z)}{dz}  &=& - 2T(z) H(z)  +  \frac{2 \eta_{\rm heat}(z)} {3 k_{\rm b}} \, \xi(z)  +   \frac{ x_{\rm ion}(z) \left [ T_\gamma(z) - T(z) \right ]}{t_{\rm c} (z)}.
\label{ion_T}
\eea
$\mu \approx 0.07$ eV$^{-1}$ is the inverse of the average ionization energy per atom, assuming 76\% H and 24\% He, neglecting double ionization of Helium \cite{arvi1}. $\alpha$ is the recombination coefficient, $T_\gamma$ is the CMB temperature, and $t_{\rm c}$ is the Compton cooling time scale $\approx$ 1.44 Myr $[30/(1+z)]^4$. The last term in the temperature evolution equation accounts for the transfer of energy between free electrons and the CMB by compton scattering \cite{coupling, recfast}. We used $x_{\rm ion} \ll 1$ and ignored the Helium number fraction in the temperature coupling term.

\subsection{Inverse compton scattering with the CMB.}

As mentioned previously, WIMP annihilation results in the production of charged particles, photons, and neutrinos. The charged particles rapidly lose energy by inverse compton scattering with CMB photons, producing a spectrum of lower energy photons of energy $E_{\rm ics}$. These ICS photons are far more efficient in ionizing and heating the gas than the prompt high energy photons \cite{hooper}. The physics of inverse compton scattering has been reviewed by \cite{ics1} and more recently by \cite{ics2}. 

At high redshifts, the energy loss is extremely rapid, with a $\sim$ 1 GeV electron losing 99\% of its energy in $\lesssim 0.02$ Myr at $z=50$, and $\lesssim 2$ years at $z=500$. The average energy of the ICS photons is given by
\beq
\langle E_{\rm ics} \rangle = \frac{4}{3} \gamma^2 \langle \epsilon \rangle = 1.6 \; \textrm{MeV} \left( \frac{E_{\rm e}(t)}{\textrm{GeV}} \right ) \; \left(  \frac{1+z}{501} \right ),
\label{energy_loss}
\eeq
where $E_{\rm ics}$ is the energy of the ICS photons, $\gamma$ is the Lorentz factor of the electron, and $\epsilon$ is the CMB energy before scattering.  We used the mean value $\langle \epsilon \rangle \approx 2.7 k_{\rm b} T_{\gamma,0} (1+z)$ in Eq. \ref{energy_loss}, assuming a black body spectrum. $\gamma$ is time dependent since the electron is losing energy.
\begin{figure}[!ht]
\begin{center}
\scalebox{0.7}{\includegraphics{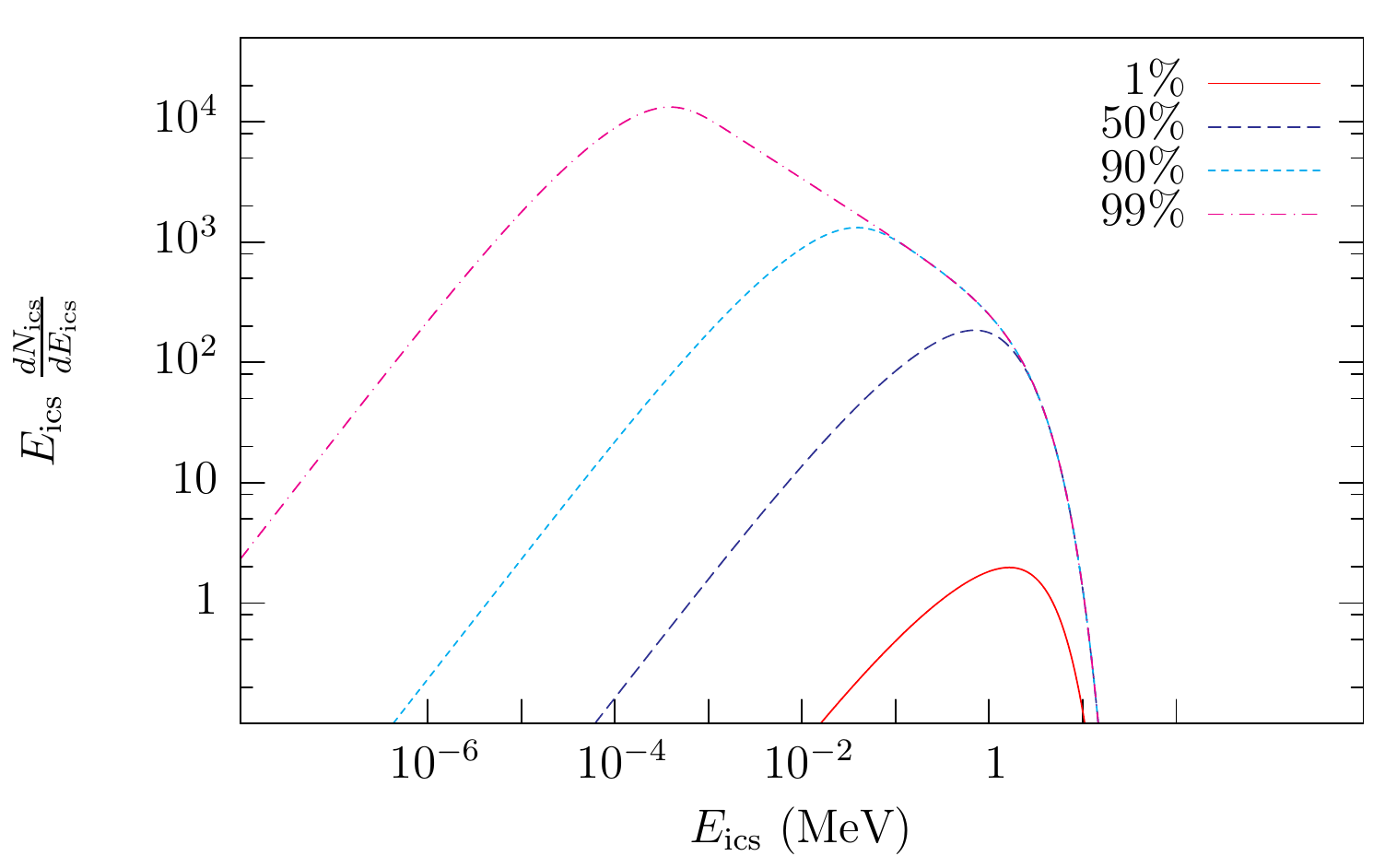}}
\end{center}
\caption{   Inverse compton scattering of a single electron of energy 1 GeV with the CMB at $z=500$. The curves show the photon spectrum at times when 1\%, 50\%, 90\%, and 99\% of the initial electron energy has been lost to the CMB.\label{fig1} }
\end{figure}

The ICS photon spectrum of a 1 GeV electron interacting with the CMB at $z=500$ is shown in Fig. \ref{fig1}. The four curves in Fig. \ref{fig1} are drawn for 1\%, 50\%, 90\%, and 99\% energy loss. Initially, the ICS photons are peaked at $\sim$ 1 MeV, in accordance with Eq. (\ref{energy_loss}). After 90\% of the initial energy is deposited into ICS photons, further scattering leads to new ICS photons with average energy $\sim (0.1)^2$ MeV. These photons add to the high energy ICS photons produced at earlier times. The result is a broad spectrum with plenty of ionizing photons. 

At high redshifts $z \sim 500$, the few $\sim$ MeV energy photons may be expected to scatter with gas atoms, and produce charged particles by pair production and compton scattering. These high energy particles will then lose energy to the CMB, adding to the ICS photon spectrum. Thus, we expect that at high redshifts, some of the high energy photons are ``recycled'' into lower energy bins. The authors of \cite{valdes} find that at low redshifts $10<z<50$, a sizeable fraction of the electron energy is converted to X-ray photons, which may free stream to the observer. The lower efficiency of inverse compton scattering at low redshifts must be taken into account when accurate results are required. In this article, we will be focused mainly on distinguishing different scenarios using CMB data, and will assume that ICS photons are of sufficiently low energy to effect ionization and heating. The detailed calculation of photon energies for arbitrary initial electron spectra is left to future work.

Assuming ionization and heating are primarily due to ICS photons with $E_{\rm ics} \lesssim$ MeV, we may make the assumption $\sigma(E_\gamma) \approx \sigma_{\rm T}$, where $\sigma_{\rm T}$ is the Thomson cross section. $\kappa$ is then energy independent, and the energy integral is simply
\beq
\int  dE_\gamma \, E_\gamma \; \frac{dN_\gamma}{dE_\gamma} (E_\gamma) = f_{\rm em} \; (2 m_\chi c^2) \; \frac{1+z}{1+z'},
\eeq 
where $f_{\rm em}$ is the fraction of energy in the form of photons/charged particles, i.e.~with electromagnetic interactions, and the energy per annihilation = $2 m_\chi c^2$.

\section{Optical depth, reionization, and CMB polarization.}

The optical depth due to scattering of microwave photons with free electrons is given by the expression
\beq
\tau(z_{\rm 1}, z_{\rm 2}) =  c \, \sigma_{\rm T} \, \int_{z_{\rm 1}}^{z_{\rm 2}}\, \frac{-dz}{H(z) (1+z)}  \, x_{\rm ion} (z) n(z) 
\eeq
$\approx 0.04$ for $z_{\rm 1} = 0, z_{\rm 2} = 6$, assuming $x_{\rm ion} = 1$, doubly ionized Helium for $z<3$, and $H(z) = H_{\rm 0} \sqrt{ \Omega_{\rm m} (1+z)^3 + \Omega_\Lambda }$. This value of $\tau(0,6)$ is less than one half of the WMAP measured optical depth. Thus, the Universe was at least partially ionized at redshifts $z>6$.  As mentioned in the Introduction, the WMAP measured $EE$ power spectrum is most sensitive to the low $l$ multipoles. The $EE$ power spectrum data at multipoles $l<10$ is sufficient when reionization is due to astrophysical objects at redshifts $z\lesssim 25$ \cite{wmap_five_year_2}. In the standard theory, this is the case. However, it is not the case when dark matter annihilation is the source of ionization. We now consider partial ionization by dark matter annihilation at intermediate, and high redshifts.

\subsection{Ionization at very high redshifts $z>100$: Annihilating free dark matter }

At redshifts $z>100$,  Eq.~(\ref{energy_equation}) may be further simplified since photons are absorbed close to the redshift at which they are released. We can approximate the exponential $e^{-\kappa}$ in Eq. (\ref{energy_equation}) as a delta function normalized so that $\mathcal{N} \int dz' e^{-\kappa} = 1$:
\beq
e^{-\kappa} \approx \frac{ (1+z) H(z) }{c \sigma_{\rm T} n(z)} \delta (z-z'),
\label{simplify}
\eeq
which makes $\xi$ independent of the cross section $\sigma$. We have verified by performing the full calculation that $\xi$ is indeed independent of $\sigma$ when $\sigma$ is close to $\sigma_{\rm T}$. Using Eq. (\ref{simplify}) and Eq. (\ref{energy_equation}), we obtain the simplified formula, valid at high redshifts $z>100$:
\bea
\xi(z) &=& f_{\rm em} \, \frac{ \langle \sigma_{\rm a} v \rangle }{m_\chi}  \; \frac{\rho^2_\chi(z)}{n(z)} \nonumber \\
&\approx& 5\times 10^{-3} \frac{ \textrm{eV} }{ \textrm{Myr} } \; \left( \frac{1+z}{501} \right )^3 \; \left( \frac{f_{\rm em}}{2/3} \right ) \frac{ \langle \sigma_{\rm a} v \rangle/(3 \times 10^{-26} \, \textrm{cm}^3/\textrm{s})}{ m_\chi / (100 \, \textrm{GeV})}
\label{simplified}
\eea
Eq. (\ref{simplified}) shows that dark matter ionization at redshifts $z>100$ can directly probe the quantity $f_{\rm em} \langle \sigma_{\rm a} v \rangle / m_\chi$, without any of the complications of the halo model.  $f_{\rm em} \approx 2/3$ assuming two thirds of the energy goes into charged particles and photons. To see the physical significance of Eq. (\ref{simplified}), let us multiply the result with a typical time scale $t_{\rm H} = 1/H(z) \approx 2.4$ Myr $\left [ 501/(1+z) \right ]^{3/2}$. Assuming 1/3 of this energy is available for ionization and heating respectively, we get $(\xi/3) t_{\rm H} \approx 4 \times 10^{-3}$ eV at $z=500$. Since the mean energy of the CMB at this redshift $= 2.7 k_{\rm b} T_{\gamma,0} (1+z) \approx 0.3$ eV, we might not expect dark matter annihilation to affect the gas or CMB temperature significantly. Assuming the same energy is available for ionization, and an ionization potential of 13.6 eV, we may expect a fraction $4 \times 10^{-3} / 13.6 \approx 3 \times 10^{-4} \left( 100 \, \textrm{GeV} /m_\chi \right )$ of the atoms to be ionized (without recombinations). Since this number is comparable to the residual electron fraction $\sim$ few $\times 10^{-4}$, we expect dark matter annihilation to contribute significantly to ionization, particularly for small $m_\chi$. We also expect the increased ionized fraction to better couple the gas and CMB temperatures by compton scattering.

Let us consider 4 different dark matter models with masses $m_\chi = 10,50,100,$ and $500$ GeV. Let $x_{\rm init}$ be the value of $x_{\rm ion}$ at our starting redshift $z=1000$.  If $x_{\rm init}$ is too small, the first term in Eq. (\ref{ion_T}) will drive $x_{\rm ion}$ up to the appropriate value. Similarly, if $x_{\rm init}$ is too large, $x_{\rm ion}$ decreases due to the second term in Eq. (\ref{ion_T}). This ensures that the value of $x_{\rm ion}$ far from the initial redshift $z=1000$ is independent of the chosen value of $x_{\rm init}$. Let  us model $x_{\rm ion}$ as a power law in $(1+z)$, valid for $z>100$, i.e.
\beq
x_{\rm ion}(z) = x_{\rm ion}(z_\ast) \, \left( \frac{1+z}{1+z_\ast} \right )^n,
\label{powerlaw}
\eeq
where $z_\ast$ is some reference redshift far from $z=1000$ where the 3 curves are nearly coincident. The exponent $n$ is determined using the value of $x_{\rm ion}$ at another redshift $z$ where the 3 curves are close to one another. We may then extend $x_{\rm ion}$ using Eq. (\ref{powerlaw}) to obtain the ``power law fit'' initial value $x_{\rm init}$. 
\begin{figure}[!h]
\begin{center}
\scalebox{0.7}{\includegraphics{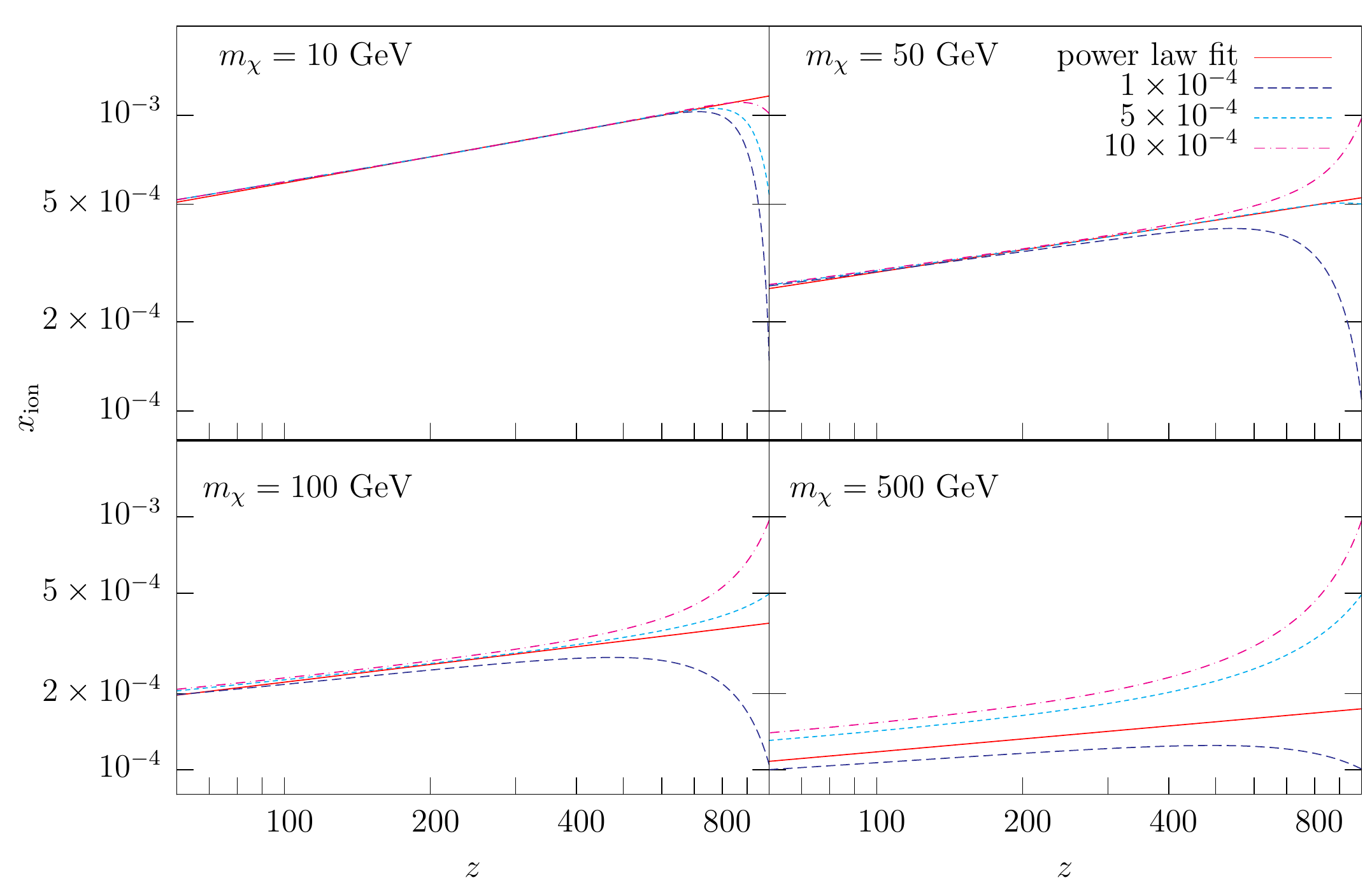}}
\end{center}
\caption{ Estimating the initial value of $x_{\rm ion}$ for the different dark matter models in the absence of residual electrons. The solid line is the power law fit, while the dashed lines are drawn for different values of $x_{\rm init}$.\label{fig2} }
\end{figure}

In Fig. \ref{fig2}, the solid(red) line denotes the power law fit to the 3 curves shown by dashed lines. For small WIMP masses, the curves become coincident quickly, while for larger masses, they approach each other only gradually. Once the value of $n$ and $[z_\ast, x_{\rm ion}(z_\ast)]$ are known for any given $m_\chi$, Eq. (\ref{powerlaw}) may be used to compute $\tau(60,1000)$, the contribution to the optical depth for $60<z<1000$. 

\subsection{Ionization at very high redshifts $z>100$: Residual electrons} 

The above discussion neglects contributions from residual electrons from the formation  
of helium and hydrogen atoms, which continues well after photons decouple (as there are many more photons than baryons/electrons in the Universe). Due to residual free electrons, 
$x_{\rm ion}$ never falls below $\sim 10^{-4}$ --- even in the absence of ionizing sources before 
$z \sim 10$.  

When residual electrons are included, the importance of the contribution to the optical depth made by dark matter annihilation is considerably reduced because the recombination term in Eq.~(\ref{ion_T}) $\propto x^2_{\rm ion}$ is dominant at high redshifts. Let $\tau_{\rm R}$ be the contribution to the optical depth made by residual electrons. To estimate $\tau_{\rm R}$, we use the {\scriptsize RECFAST} code described in \cite{recfast}, and implemented in the {\scriptsize CAMB} program \cite{camb}.  The contribution of residual electrons to the optical depth has been studied by \cite{shull_venkatesan}, who find that residual electrons may contribute $\approx 0.06$ to the optical depth between redshifts $7<z<700$. The optical depth contribution $\tau_{\rm R}(z)$ increases sharply for $z>700$. Table 1 lists the values of  $\tau(60,1000)$ obtained using the power law fit of $x_{\rm ion}$ for  WIMP masses $m_\chi = 10,50,100$, and 500 GeV respectively, with and without residual electrons. As seen from Table 1, the effect of residual electrons is to reduce the impact of the optical depth contribution of dark matter annihilation to low values except for the case of very light particle masses, i.e.~dark matter annihilation is ineffective when the residual electron fraction $\gg \xi(z)/[(13.6\; {\rm eV}) H(z)]$. Nevertheless, dark matter annihilation always increases the total optical depth. Thus, one needs to carefully consider the recombination history 
when placing constraints on dark matter annihilation at high redshifts.

\begin{table} [!h] {Values of $\tau (60,1000)$ for different dark matter masses.   \\ }
\vspace{0.1in}
\begin{tabular}  {|c| |c |c |c |c |} 
   \hline 
   $m_\chi$ (GeV) & 10 & 50 & 100 & 500 \\  \hline  
   $\tau(60,1000)$ ($\tau_{\rm R} = 0$) & 0.070 & 0.033 & 0.024 & 0.011\\  \hline 
   $\tau(60,1000) - \tau_{\rm R}$ & 0.028 & 0.007 & 0.004 & 0.001\\      \hline
   $\tau(60,1000)$ & 0.429 & 0.407 & 0.404 & 0.401\\      \hline
   $n$ & 0.295 & 0.253 & 0.234 & 0.171\\      \hline
   $x_{\rm ion}(z_\ast) \times 10^4$ & 8.9 & 4.2 & 3.1 & 1.5\\               
   \hline   
\end{tabular}

\vspace{0.1in}
\caption{ Estimated values of the optical depth for the different dark matter models with and without residual electrons. The row $\tau(60,1000) (\tau_{\rm R}=0)$ is due to dark matter annihilation only, and assumes zero residual electron contribution. The next two rows include residual electrons computed using the {\scriptsize RECFAST} code.  $\tau(60,1000) - \tau_{\rm R}$ is the contribution due to dark matter in the presence of residual electrons, while $\tau(60,1000)$ is the full optical depth, which includes contributions due to both dark matter and residual electrons. Most of the contribution to $\tau(60,1000)$ comes from $z>700$. $n$ and $x_{\rm ion}(z_\ast)$ are defined in the text. $z_\ast \approx 400$. The WMAP measured value of optical depth is mostly due to ionization at late times ($z \lesssim 25$) and does not include the contribution due to residual electrons. }  
\end{table}

\subsection{Reionization and CMB power spectra.}

We now consider dark matter models at both high and intermediate redshifts, and compare these with a standard, two step baryonic reionization model (inspired by simulations as e.g.~\cite{Gnedin:1996qr}), referred to as Model A. We use the term `baryonic' simply to indicate that reionization is entirely due to luminous astrophysical sources, as opposed to dark matter annihilation. We consider a 2-step reionization scenario for Model A.  From $z=25$ to $z=6$, we assume a linear increase of $x_{\rm ion}(z)$ with $z$, with full ionization for $z<6$. $x_{\rm ion}(z)$ is smoothly patched on to the value of $x_{\rm ion}$ for $z<6$ and $z>25$  using a fitting function of the hyperbolic tangent type. The particular form of $x_{\rm ion}$, as well as the fitting function for this model are chosen primarily for convenience, and are not motivated by any particular theory of star formation. 

We also consider 3 models motivated by dark matter annihilation. In these models, reionization is due to both dark matter and luminous sources. For $z<25$, we assume the same reionization scenario considered in Model A. For intermediate redshifts $25<z<60$, partial reionization is due to dark matter annihilation in halos. For $z>60$, free dark matter particles play the dominant role. In the intermediate redshift range $25<z<60$, where dark matter halos are the primary source of ionization, we assume a Navarro-Frenk-White \cite{nfw} type density profile, and a Sheth-Tormen \cite{st} type distribution of halo masses. The concentration parameter of the halos is treated as a free parameter. DM Model \#1 has a particle mass $m_\chi = 100$ GeV, and a concentration parameter $c = 15$. DM Model \#2 has $m_\chi$ = 10 GeV, $c$ = 5, and DM Model \#3 has $m_\chi$ = 10 GeV, $c$ = 15. The optical depth at low redshifts $\tau(0,25)$ is set to a value of 0.087 for all models, consistent with the measurements reported by WMAP.

\begin{figure}[!h]
\begin{center}
\scalebox{0.7}{\includegraphics{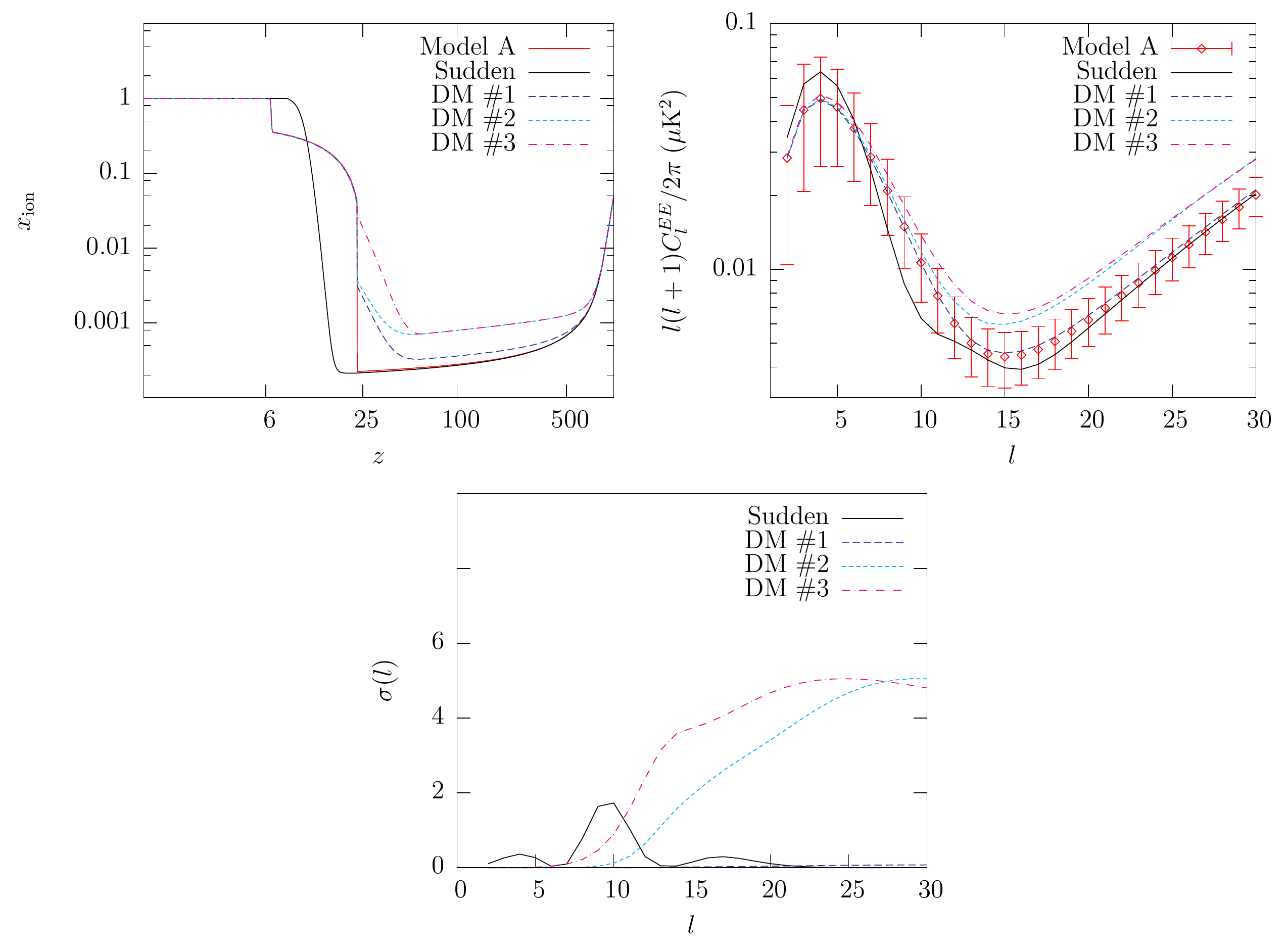}}
\end{center}
\caption{  
Comparing the 3 dark matter models with baryonic Model A. For $z<25$, the dark matter models are identical to Model A, and have an optical depth $\tau(0,25) = 0.087$. For $z>25$, the dark matter models feature partial ionization due to halos and free particles. Also shown for comparison is a `sudden' reionization model. (a) shows the ionization history for the different models. (b) shows the respective polarization power spectra ($EE$), with $1\sigma$ cosmic variance error bars computed using the power spectrum of Model A. (c) shows the $\sigma(l)$ deviations of the different models ($EE$ power spectra), all compared to Model A. The solid (black) curve denotes the `sudden' reionization model, while the 3 dashed curves are drawn for the 3 dark matter models. DM Model \#1 is indistinguishable from 
Model A. \label{fig3} }
\end{figure}

To quantify the difference between two models $X$ and $Y$, let us compute the $\sigma(l)$ deviation for the $E$ mode polarization, defined as:
\beq
\sigma^2(l) =  \left( \frac{ C^{X}_l - C^{Y}_l }{\Delta C_l} \right )^2. 
\label{sig}
\eeq
$C_l = C^{EE}_l$ is the 2 point $EE$ polarization power spectrum for multipole $l$. $\Delta C_l = \Delta C^{EE}_l$ is the $1\sigma$ cosmic variance error at $l$: 
\beq
\left( \Delta C_l \right )^2 =  \frac{2}{2l+1}  \; C^2_l,
\label{error_bar}
\eeq
which neglects instrument noise and assumes 100\% sky coverage. In general, $C^2_l$ in Eq. (\ref{error_bar}) should be replaced by
\beq
\left( C^{AB}_l  \right )^2 = \frac{1}{2} \left [ C^{AA}_l C^{BB}_l  + \left( C^{AB}_l \right )^2 \right ]
\eeq
where $A$ and $B$ could stand for $T$ or $E$. Instrument noise is not negligible in real experiments. For example, instrument noise for the Planck experiment is expected to become comparable to cosmic variance at $l=20$ \cite{planck}. However, in this article, we use Eq. (\ref{error_bar}) and consider a cosmic variance limited observation.

Fig. \ref{fig3}(a) shows the ionization histories of the different models. The baryonic model A is shown by the solid (red) curve. For comparison, we also include a `sudden' reionization model, shown by the solid (black) curve. In the sudden reionization model, the Universe is ionized below $z\sim11$.  The three dark matter models are shown using dashed lines. We note that the two models with $m_\chi = 10$ GeV (DM Model \#2 and DM Model \#3) have larger $x_{\rm ion}$ at high redshifts, compared to the model with $m_\chi = 100$ GeV (DM Model \#1). Also of note is the effect of the concentration parameter $c$. (b) shows the respective polarization power spectra.  The error bars are cosmic variance only, and are calculated using Eq (\ref{error_bar}) for model A.  The dashed curves are drawn for the 3 dark matter models. (c) shows the $\sigma(l)$ variation (comparing $EE$ power spectra) between the 3 dark matter models and Model A. The solid (black) curve denotes the sudden reionization model compared with Model A. The difference between DM Model \#1 and Model A is nearly zero. DM Models \#2 and \#3 show a larger deviation and may be distinguished from Model A at high confidence with cosmic variance limited data up to $l=20$. With cosmic variance limited data up to $l=10$, the dark matter models are not easily distinguishable from the standard reionization scenario. Considering that WMAP polarization data is cosmic variance limited only for $2<l<6$, and the signal is not significant for $l>10$, current large angle polarization observations cannot distinguish between different reionization histories, but future experiments may be able to do better. \cite{cmbpol} have considered separately, ionization at redshifts $6<z<10$ and $z>10$ and have constructed exclusion contours given the WMAP data, as well as predictions for Planck and CMBPol. However, the fiducial ionization history considered in \cite{cmbpol} does not include ionization by dark matter at redshifts $z>100$. We conclude that in the absence of polarization data for $l>10$, it is hard to constrain dark matter annihilation ($m_\chi > 10$ GeV, thermal relic cross section), or differentiate between different reionization models using large angle CMB polarization.

\begin{figure}[!h]
\begin{center}
\scalebox{0.4}{\includegraphics{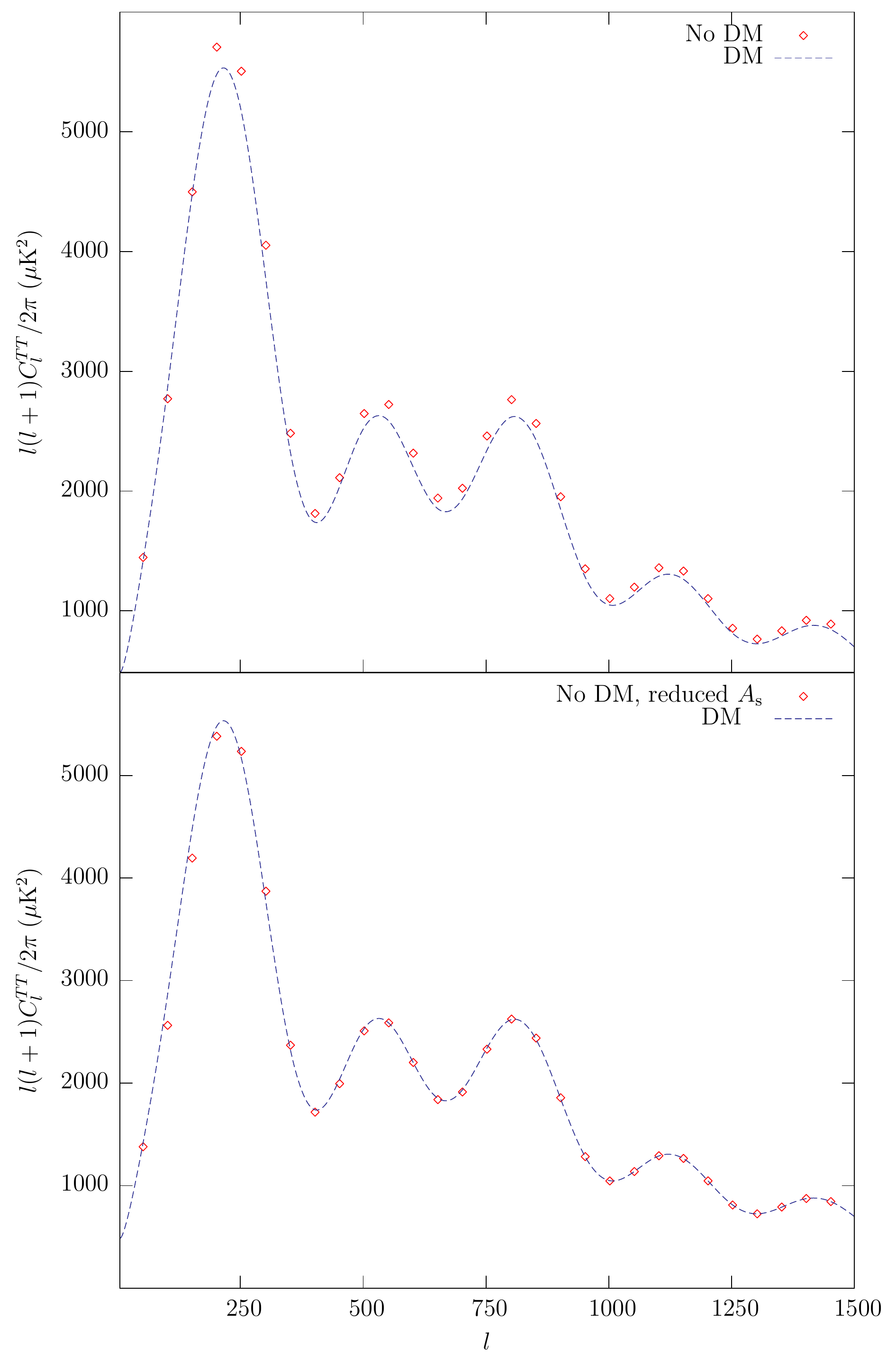}}
\end{center}
\caption{The $TT$ power spectrum. Partial ionization by dark matter results in a larger optical depth, causing a suppression of power (top panel). Decreasing the amplitude of the primordial scalar power $A_{\rm s}$ by $\sim 5\%$ can mimic the effect of $\tau(z)$) (bottom panel).  \label{fig4} }
\end{figure}

\begin{figure}[!h]
\begin{center}
\scalebox{0.4}{\includegraphics{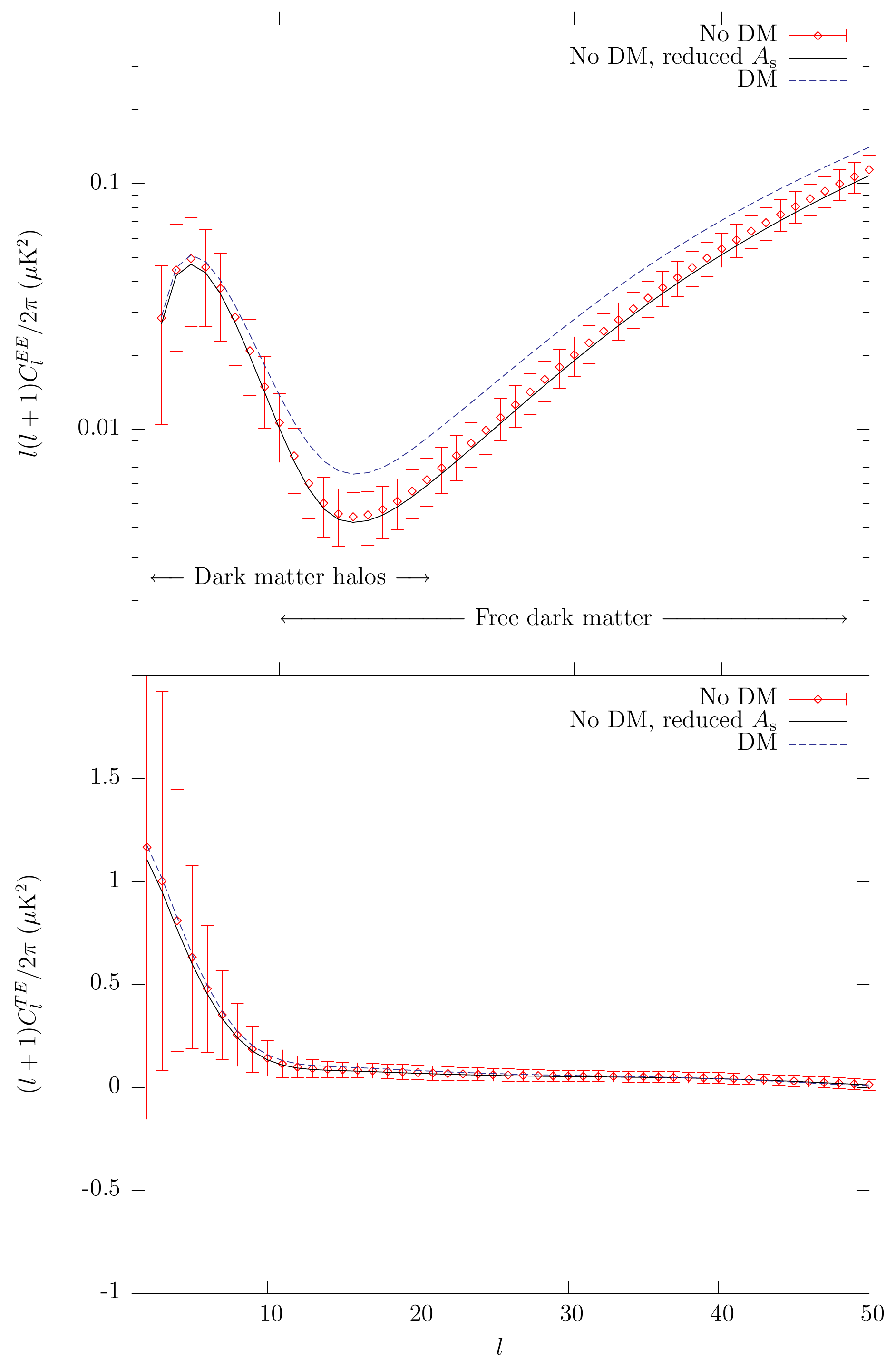}}
\end{center}
\caption{The large angle $EE$ (top panel) and $TE$ (bottom panel) power spectra. The $1\sigma$ cosmic variance error bars are drawn for baryonic Model A. The solid (black) curve is drawn for the baryonic model with reduced $A_{\rm s}$ . The dashed curve (blue) is plotted for dark matter Model \#3 ($m_\chi$=10 GeV, $c$=15).  It is harder to distinguish different models using the $TE$ power spectrum because $C^{TE}_l \propto \tau(z)$, while $C^{EE}_l \propto \tau(z)^2$.  Following the WMAP convention, the bottom panel shows $(l+1)C^{TE}_l/2\pi$ and not $l(l+1)C^{TE}_l/2\pi$ \label{fig5} }
\end{figure}

One may consider the possibility of using the temperature power spectrum to tighten constraints on 
$\tau(z)$. The $TT$ power spectrum is damped by a factor $\sim e^{-2\tau(z)}$ on sufficiently small 
scales due to free electrons scattering with microwave photons. The WMAP $TT$ power spectrum is cosmic variance limited for $l<548$ \cite{wmap_7year_1} and thus, the effect of dark matter annihilation 
is potentially observable in the $TT$ spectrum. The top panel of Fig. \ref{fig4} shows the $TT$ power spectrum (red points) with no dark matter annihilation taken into account, compared with dark matter Model \#3 ($m_\chi = 10$ GeV, $c$ = 15) shown by dashed lines. As expected, the power is slightly reduced when dark matter annihilation is included, due to the larger optical depth. However, this damping is nearly indistinguishable from the effect of reducing the overall normalization of the power spectrum set by $A_{\rm s}$, the amplitude of the primordial scalar power at the end of inflation. The bottom panel of Fig.\ref{fig4} compares the $TT$ power spectrum with $A_s$ lowered by $\sim 5\%$ to the dark matter model, showing very good agreement. Reducing $A_{\rm s}$ would result in slightly lower power on large scales in the $TT$ spectrum (which are unaffected by reionization at high redshifts), but this difference is hindered by the larger cosmic variance at small $l$. 

The correlation between $\tau(z)$ and $A_{\rm s}$ may be quantified by means of the correlation parameter $\rho$ constructed from the inverse of the Fisher matrix. For a $2 \times 2$ Fisher matrix, one readily finds:
\beq
\rho = - \frac{F_{ij}}{\sqrt{F_{ii} F_{jj}}}
\eeq
where 
\beq
F_{ij} = \sum_l \frac{C_{l,i}}{\delta C_l} \; \frac{C_{l,j}}{\delta C_l}.
\eeq
$\delta C_l$ represents the $1\sigma$ error in $C_l$ and the commas indicate a derivative w.r.t the parameter being varied. For an assumed form of $C_l \sim A_{\rm s} e^{-2\tau(z)}$, we find $\rho = 1$, implying the Fisher matrix is singular, with infinitely large errors in $\tau(z)$ and $A_{\rm s}$. In practice, $\rho$ is not exactly 1 and the Fisher matrix is invertible.
One might also ask whether alternate scenarios may be distinguishable from the effect of dark matter annihilation. To test this, we constructed a reionization history by modifying the recombination coefficient as follows:
\beq
 \alpha_{\rm new}(z) = \alpha(z) \left[ \frac{1+z}{1+z_{\rm 0}} \right ]^p
 \label{mod}
 \eeq
at redshifts below $z_{\rm 0} \approx 800$, i.e. when the residual electrons start to become sub dominant. $\alpha(z)$ is the standard recombination coefficient (see Eq. (\ref{ion_T})). Thus, Eq. (\ref{mod}) considers an ionization history with a faster fall off of the recombination coefficient with redshift. We calculate $\rho$ by considering a simple 3 parameter Fisher matrix consisting of $\tau(z)$, $A_{\rm s}$ and spectral index $n_{\rm s}$, and marginalizing over $n_{\rm s}$.  More parameters need to be included for a careful study, but this simple analysis suffices to show the degree of correlation between the parameters. We consider two cases with excess optical depth: (i) dark matter annihilation at high redshifts (with $m_\chi$ = 50 GeV, no halos) with the standard recombination coefficient, and (ii) no dark matter, but a smaller recombination coefficient as in Eq. (\ref{mod}) (with $p$ = 0.985). For the $TT$ power spectrum, we find for case (i), a correlation between $\tau(z)$ and $A_{\rm s}$ of $\rho \approx 0.9989$. For case (ii), we find $\rho \approx 0.9993$. The $1\sigma$ errors in $\tau(z)$ given by $\sqrt{ [{F^{-1}}]_{11}}$ are found to be $\approx 0.018$ for case (i) and $\approx 0.023$ for case (ii). The exact values of $\rho$ and $\delta \tau(z)$ will be different when more parameters are included. However, this simple calculation shows that $\tau(z)$ and $A_s$ are strongly degenerate. We also see that the effects of dark matter annihilation cannot easily be distinguished from an alternate recombination theory.

Fig. \ref{fig5} shows the corresponding polarization power spectra. We verify that reducing $A_{\rm s}$ has almost no effect on the large angle polarization power. Thus, while the dark matter model cannot be separated from the standard scenario using the $TT$ power spectrum when $A_{\rm s}$ is variable, they are easily separable using $EE$ polarization data provided instrument noise is not a limiting factor. Also shown is the large angle $TE$ power spectrum. One sees that the $EE$ power spectrum is more useful than the $TE$ power spectrum for distinguishing models, provided quality $EE$ data exists for $l\lesssim 20$. The contribution due to dark matter halos mostly affects multipoles $l\lesssim 20$, while annihilation of free dark matter at high redshifts influences the multipoles $l>10$. An ideal, cosmic variance limited observation would be able to place constraints on both the halo model, as well as particle physics parameters. The Planck experiment is expected to be cosmic variance limited up to $l\sim10$ \cite{cmbpol} and have good signal to noise up to $l\sim 20$ \cite{planck}. The CMBPol experiment is expected to perform even better. These future experiments may be able to distinguish between different reionization models, and place constraints on dark matter halos as well as particle parameters.

\section{Conclusions.}

In this article, we have looked at how dark matter models influence the CMB polarization power spectrum, and whether current, and future observations can be used to distinguish dark matter reionization models from the standard theory. 

In Section II, we briefly reviewed dark matter 
annihilation resulting in ionization and heating, and derived an expression for the energy absorbed per gas atom $\xi(z)$. We also considered inverse compton scattering of high energy charged particles 
with the CMB, resulting in lower energy photons (Fig. 1). When the inverse compton process is efficient, $\xi(z)$ can be expressed in a much simpler form. At very high redshifts, $\xi(z)$ can be further 
simplified since most of the released energy is absorbed close to the redshift of particle annihilation. In this regime, the ionization directly probes $f_{\rm em} \langle \sigma_{\rm a} v \rangle / m_\chi$, where 
$f_{\rm em}$ is the fraction of energy transported by electromagnetic processes. In Section III, we discussed the optical depth contribution at different redshifts and the resulting CMB polarization power spectrum. We first considered dark matter annihilation at high redshifts $z>100$, both with and without residual electrons (Fig. 2, Table 1). Taking residual electrons into account makes dark matter annihilation less effective except for very small particle masses. We then considered dark matter annihilation at all redshifts, and calculated the ionization history for different dark matter masses, and concentration parameters (Fig. 3). We also considered the temperature power spectrum, and showed that it is not as useful as the polarization power spectrum in distinguishing different reionization scenarios (Figs 4,5). The $EE$ power spectrum is more sensitive to reionization than the $TE$ power spectrum.

In this article, we have restricted ourselves to the case of a thermal relic with cross section $\langle \sigma_{\rm a} v \rangle = 3 \times 10^{-26}$ cm$^3/$s. However, dark matter annihilation at high redshifts probes $f_{\rm em}  \langle \sigma_{\rm a} v \rangle / m_\chi$ and thus, our results are valid for a wide range of masses, for corresponding annihilation cross sections. We considered the effect of different particle masses, as well as different concentration parameters. The halo contribution which is dominant for $z<60$ affects the low multipoles $l \lesssim 20$ in the polarization power spectrum, while free dark matter annihilation at redshifts $z>60$ primarily affects multipoles $l>10$. We first looked at DM Model \#1 with $m_\chi=100$ GeV and $c=15$, and compared it with a 2-step standard reionization model (Model A). The effect of dark matter is not seen in the large angle polarization, even with cosmic variance limited data up to $l=30$. Dark matter annihilation becomes important when the particle mass is decreased. We considered two models with $m_\chi = 10$ GeV, with concentration parameters $c=5$ and $c=15$ (DM Model \#2, and DM Model \#3 respectively). With cosmic variance limited data up to $l \sim20$, DM Model \#2 and DM Model \#3 may be distinguished at high significance from Model A.  
Thus future CMB polarization measurements might allow us to put constraints on light (on the weak energy scale) dark matter candidates and on models with an enhanced annihilation cross section. Current large angle CMB polarization measurements are consistent with a standard baryonic reionization theory. Future experiments such as Planck would be able to shed more light on the physics of reionization.

\acknowledgments{We thank Nishikanta Khandai, Roberto Trotta, Licia Verde, and Eiichiro Komatsu for helpful discussions.  A.N. thanks the Bruce and Astrid McWilliams Center for Cosmology for financial support.  D.S. was supported by the Deutsche Forschungsgemeinschaft under grant GRK 881. }

\end{document}